\documentclass[preprint2]{aastex}
\usepackage{graphicx}
\usepackage{color}

\begin{document}

\title{On the search for artificial Dyson-like structures around pulsars}

\author{Z. Osmanov}
\affil{School of Physics, Free University of Tbilisi, 0183, Tbilisi,
Georgia}

\begin{abstract}
Assuming the possibility of existence of a supercivilization we
extend the idea of Freeman Dyson considering pulsars instead of
stars. It is shown that instead of a spherical shell the
supercivilization must use ring-like constructions. We have found
that a size of the "ring" should be of the order of
$(10^{-4}-10^{-1})$AU with temperature interval $(300-600)$K for
relatively slowly rotating pulsars and $(10-350)$AU with temperature
interval $(300-700)$K for rapidly spinning neutron stars,
respectively. Although for the latter the Dyson construction is
unrealistically massive and cannot be considered seriously.
Analyzing the stresses in terms of the radiation and wind flows it
has been argued that they cannot significantly affect the ring
construction. On the other hand, the ring in-plane unstable equilibrium
can be restored by the energy which is small compared to the energy extracted from
the star. This indicates that the search for infrared ring-like
sources close to slowly rotating pulsars seems to be quite
promising.

\end{abstract}


%%%%%%%%%%%%%%%%%%%%%%%%%%%%%%%%%%%%%%
\section{Introduction}
%%%%%%%%%%%%%%%%%%%%%%%%%%%%%%%%%%%%%%

In the framework of SETI (Search for Extraterrestrial Intelligence)
one of the important projects was the search for the interstellar
radio communication. A rather different and quite original method
was proposed by the prominent physicist Freeman Dyson in 1960, who
suggested that if an extraterrestrial intelligence has reached a
level of supercivilization, it might consume an energy of its own
star \citep{dyson}. For this purpose, to have the maximum efficiency
of energy transformation, it would be better to construct a thin
shell completely surrounding the star. In the framework of this
approach the author assumes that the mentioned superintelligence
observed by us will have been in existence for millions of years,
having reached a technological level exceeding ours by many orders
of magnitude. Kardashev in his famous article \citep{kardashev},
examining the problem of transmission of information by
extraterrestrial civilizations, has classified them by a
technological level they have already achieved: (I) - a
technological level close to the level of the civilization on earth,
consuming the energy of the order of $4\times 10^{19}$ergs s$^{-1}$;
(II) - a civilization consuming the energy of its own star -
$4\times 10^{33}$ergs s$^{-1}$ and (III) - a civilization capable of
harnessing the energy accumulated in its own galaxy: $4\times
10^{44}$ergs s$^{-1}$. In this classification Dyson's idea deals
with the civilization of type-II, consuming the energy exceeding
ours approximately $\frac{4\times 10^{33}}{4\times 10^{19}}=10^{14}$
times (see the similar estimates by \cite{dyson}). If we assume that
an average growth rate of $1\%$ per year in industry is maintained,
the level of type-II civilization might be achieved in $\sim 3000$
years \citep{dyson}, being quite reasonable in the context of the
assumption that a civilization exists millions of years.

%\begin{figure}
%  \centering {\includegraphics[width=7cm]{test4.eps}}
%  \caption{On the picture we schematically show the pulsar, its axis of rotation,
%  and two emission channels with an opening angle $\beta$. It is worth noting that
%  when $\alpha$ is close to $90^o$, the Dyson construction has to be located in the equatorial
%  plane. Contrary to this, for relatively smaller angles, the emission channels
%  will irradiate two different ring-like structures located in different planes
%  parallel to that of the equator.}\label{fig1}
%\end{figure}

Dyson has suggested that if such a civilization exists, then it is
possible to detect it by observing the spherical shell surrounding
the star. In particular, it is clear that energy radiated by the
star must be absorbed by the inner surface of the sphere and might
be consumed by the civilization. The author implied that the size of
the sphere should be comparable with that of the Earth's orbit. It
is clear that to have energy balance, the spherical shell must
irradiate the energy, but in a different spectral interval: in the
infrared domain, with the black body temperature of the order of
$(200-300)$K \citep{dyson}.

The attempts to identify Dyson spheres on the sky were performed by
several groups \cite{jugaku,slish,timofeev} but no significant
results were obtained. Recently an interest to such an ambitious
idea has significantly increased: a couple of years ago Carrigan has
published an article titled: "IRAS-based whole sky upper limit on
Dyson spheres" \citep{carrigan}, where he considered the results of
the instrument IRAS (The Infrared Astronomical Satellite). This
satellite covered almost $96\%$ of the sky, implying that this is
almost whole sky monitoring. According to the study, the searches
have been conducted as for fully as for partially cloaked Dyson
spheres. The search has revealed $16$ Dyson sphere candidates, but
the author pointed out that further investigation must have been
required. Recently an interesting series of works has been presented
\citep{wright1,wright2,wright3} where
the authors discuss a broad class of related problems starting from
philosophical aspects of SETI \citep{wright1}, also examining in detail the strategy of the
search for the infrared galactic and extragalactic sources corresponding to
Kardashev-II/III civilizations \citep{wright2,wright3}.

\begin{figure}
  \centering {\includegraphics[width=7cm]{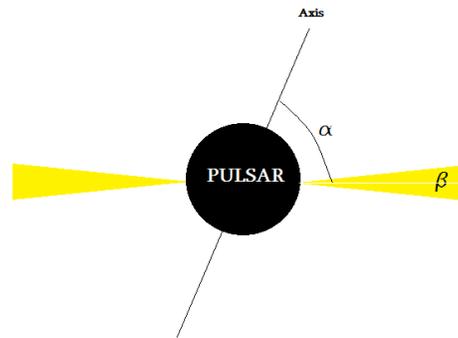}}
  \caption{On the picture we schematically show the pulsar, its axis of rotation,
  and two emission channels with an opening angle $\beta$. It is worth noting that
  when $\alpha$ is close to $90^o$, the Dyson construction has to be located in the equatorial
  plane. Contrary to this, for relatively smaller angles, the emission channels
  will irradiate two different ring-like structures located in different planes
  parallel to that of the equator.}\label{fig1}
\end{figure}

In this manuscript we present a rather different idea how to search
an advanced intelligence. In the framework of Dyson's idea, the
spherical shroud (with radius of the order of $(1-3)$AU) is
constructed around stars. It is clear that in order to consume
almost the total energy radiated by the star, it must be imbedded
inside a closed spherical shell, requiring enormous material to
construct it. On the other hand, it is very well known that pulsars
- rapidly rotating neutron stars - emit huge energy in narrow
channels (see Figure \ref{fig1}), therefore, if a supercivilization
exists it can in principle utilize the energy of these objects. But
in these cases, instead of sphere-like envelopes the super
intelligence has to use ring-like structures around the pulsars. If
the angle between the rotation axis and the direction of emission,
$\alpha$, is close to $90^o$, the ring will be located in the
equatorial plane of the neutron star. As we will see later, in case
of relatively slowly spinning pulsars (with periods of rotation of
the order of $1$s) an advantage will be quite small sizes of these
artificial constructions. Another class of neutron stars is the
so-called $X$-ray pulsars, having a strong emission pattern in the
$X$-ray band. Usually these objects are characterized by short
periods of rotation with quite high values of slowdown rates, having
luminosities exceeding those of normal pulsars by several orders of
magnitude. It is worth noting that a habitable zone around them
might be further than around slowly spinning pulsars,
implying that the possible size of the artificial "ring" could be
extremely large. Therefore, we can hardly believe that these objects
can be interesting in the context of the search for extraterrestrial
super advanced intelligence.

The organization of the paper is the following: after introducing
the theoretical background we work out the details of the Dyson
"rings" surrounding the pulsars in Sec.2, in Sec. 3 we summarize our
results.

%%%%%%%%%%%%%%%%%%%%%%%%%%%%%%%%%%%%%%%%%%%%%%%
\section[]{Theoretical background and results}
%%%%%%%%%%%%%%%%%%%%%%%%%%%%%%%%%%%%%%%%%%%%%%%

In this section we consider the pulsars and estimate the
corresponding physical parameters of artificial ring-like
constructions surrounding the rotating neutron stars.

Generally speaking, any rotating neutron star, characterized by the
slow down rate $\dot{P}\equiv dP/dt>0$, where $P$ is the rotation
period, loses energy with the following power (called the slowdown
luminosity, $L_{sd}$) $\dot{W} = I\Omega|\dot{\Omega}|$. Here by
$I=2MR_{\star}^2/5$ we denote the moment of inertia of the neutron
star, $M\approx 1.5\times M_{\odot}$ and $M_{\odot}\approx 2\times
10^{33}$g are the pulsar's mass and the solar mass respectively,
$R_{\star}\approx 10^6$cm is the neutron star's radius,
$\Omega\equiv 2\pi/P$ is its angular velocity and
$\dot{\Omega}\equiv d\Omega/dt=-2\pi\dot{P}/P^2$. The slowdown
luminosity for the relatively slowly spinning pulsars is of the
order of
$$L_{sd}\approx 4.7\times 10^{31}\times
P^{-3}\times\left(\frac{\dot{P}}{10^{-15}ss^{-1}} \right)\times$$
\begin{equation}
\label{lumin} \times\left(\frac{M}{1.5M_{\odot}}\right)ergs\ s^{-1},
\end{equation}
where the parameters are normalized on their typical values. As it
is clear from Eq. (\ref{lumin}), the energy budget is very high
forcing a supercivilization construct a "ring" close to the host
object. On the other hand, these sources exist during the time scale $P/\dot{P}$,
which is long enough to colonize the star.

In the framework of the standard definition, a habitable zone (HZ)
is a region of space with favorable conditions for life based on
complex carbon compounds and on availability of fluid water, etc.
\citep{hanslmeier}. This means that the surface of the "ring" must
be irradiated by the same flux as the surface of the Earth
(henceforth - the flux method). Therefore, the mean radius of the
HZ, defined as $R_{HZ} = \left(\kappa
L_{sd}/L_{\odot}\right)^{1/2}$AU, where $L_{\odot}\approx 3.83\times
10^{33}$ergs s$^{-1}$ is the solar bolometric luminosity, is
expressed as follows
$$R_{_{HZ}}\approx 3.5\times 10^{-2}\times
P^{-3/2}\times\left(\frac{\kappa}{0.1}\right)^{1/2}\times$$
\begin{equation}
\label{rHZ} \times\left(\frac{\dot{P}}{10^{-15}ss^{-1}}
\right)^{1/2}\times\left(\frac{M}{1.5M_{\odot}}\right)^{1/2}AU,
\end{equation}
where we have taken into account that the bolometric luminosity,
$L$, of the pulsar is less than the slowdown luminosity and is
expressed as $\kappa L_{sd}$, $\kappa < 1$. As we see, the radius of
the HZ for $1$s pulsars is very small compared to the radius of the
typical Dyson sphere - $1$AU. The best option for the
supercivilization could be to find a pulsar with an angle between
the magnetic moment (direction of one of the channels) and the axis
of rotation close to $90^o$, because in this case an artificial
"ring" has to be constructed in the equatorial plane. In case of
relatively small inclination angles, there should be two rings, each
of them shifted from the equatorial plane, although, in this case it
is unclear how do the "rings" keep staying in their planes,
therefore, we focus only on pulsars with $\alpha\approx 90^o$.

According to the standard model of pulsars it is believed that the
radio emission is generated by means of the curvature radiation
defining the opening angle of the emission channel \citep{rudsuth}
$$\beta\approx 32^o\times
P^{-13/21}\times\left(\frac{\rho}{10^6cm}\right)^{2/21} \times$$
\begin{equation}
\label{angle} \times\left(\frac{\dot{P}}{10^{-15}ss^{-1}}
\right)^{1/14}\times\left(\frac{\omega} {1.0\times
10^{10}Hz}\right)^{-1/3},
\end{equation}
where the radius of curvature of magnetic field lines - $\rho$ and
the cyclic frequency of radio waves - $\omega$, are normalized on
their typical values \citep{rudsuth}. One can straightforwardly
check that the artificial "ring" with equal radiuses of the
spherical segment bases should have height in the following interval
$\left(1-2\right)R_{_{HZ}}\sin\left(\beta\right)\approx\left(0.006-0.06\right)$AU.

One of the significant parameters to search the "rings" is their
temperature, which can be estimated quite straightforwardly. In
particular, if the aim of the super civilization is to consume the
total energy radiated by the host pulsar, then albedo of the
material the "ring" has to be made of must be close to zero. For
simplicity we consider $\alpha = 90^o$, then it is clear that the
inner surface of the "ring", irradiated by the pulsar, every single
second absorbs energy of the order of $L$. On the other hand, energy
balance requires that the "ring" must emit the same amount of energy
in the same interval of time, $L = A_{ef}\sigma T^4$, leading to the
following expression
\begin{equation}
\label{balance} T = \left(\frac{L}{A_{_{ef}}\sigma}\right)^{1/4},
\end{equation}
where $\sigma\approx 5.67\times 10^{-5}$erg/(cm$^2$K$^4$) is the
Stefan-Boltzmann constant, $A_{_{ef}} = 8\pi
R_{_{HZ}}^2\sin\left(\alpha/2\right)$ is the effective area of the
spherical segment taking into account inner and outer surfaces and
$T$ is the average temperature of the "ring".

%\begin{figure}
%\resizebox{\hsize}{!}{\includegraphics{fig1a.eps}}
%\resizebox{\hsize}{!}{\includegraphics{fig1b.eps}}
% \caption{On the top panel, in the framework of
%the flux method, we plot the dependence of $T$ on $P$ for three
%different values of $\dot{P}$: $\dot{P} = 10^{-15}$ss$^{-1}$ (solid
%line); $\dot{P} = 10^{-14}$ss$^{-1}$ (dashed line); $\dot{P} =
%2\times 10^{-14}$ss$^{-1}$ (dotted-dashed line). As it is clear from
%the graph, for typical values of relatively slowly spinning pulsars
%the effective temperature of the artificial construction ranges from
%$\sim 400$K to $\sim 500$K. On the bottom panel, in the framework of
%the temperature method, we show the dependence of $R_{HZ}$ on $T$
%for the same values of $\dot{P}$. As we see the distance to the HZ
%ranges from $2\times 10^{-4}$AU to $1.3\times 10^{-3}$AU.}
%\label{fig2}
%\end{figure}

\begin{figure}
\resizebox{\hsize}{!}{\includegraphics{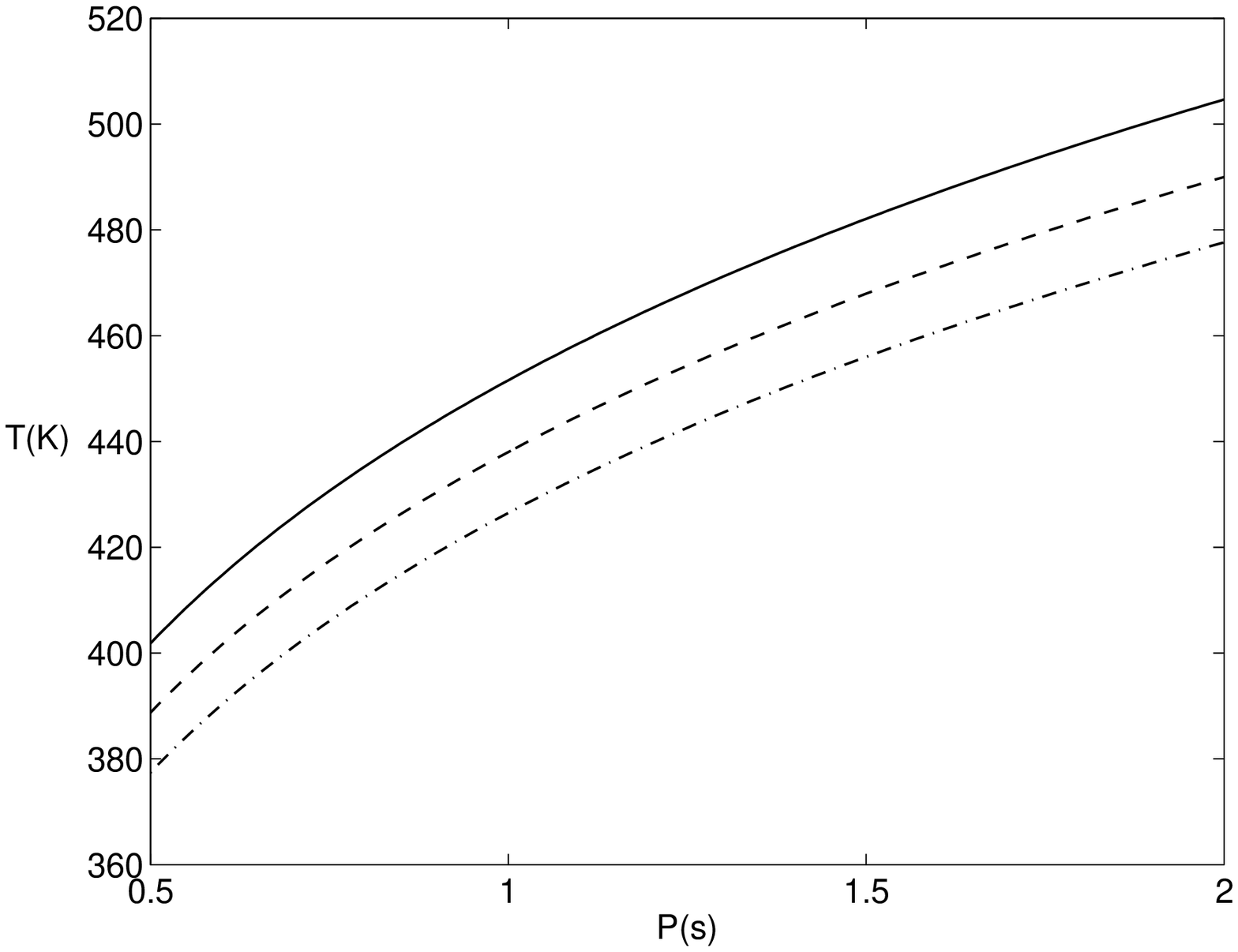}}
\resizebox{\hsize}{!}{\includegraphics{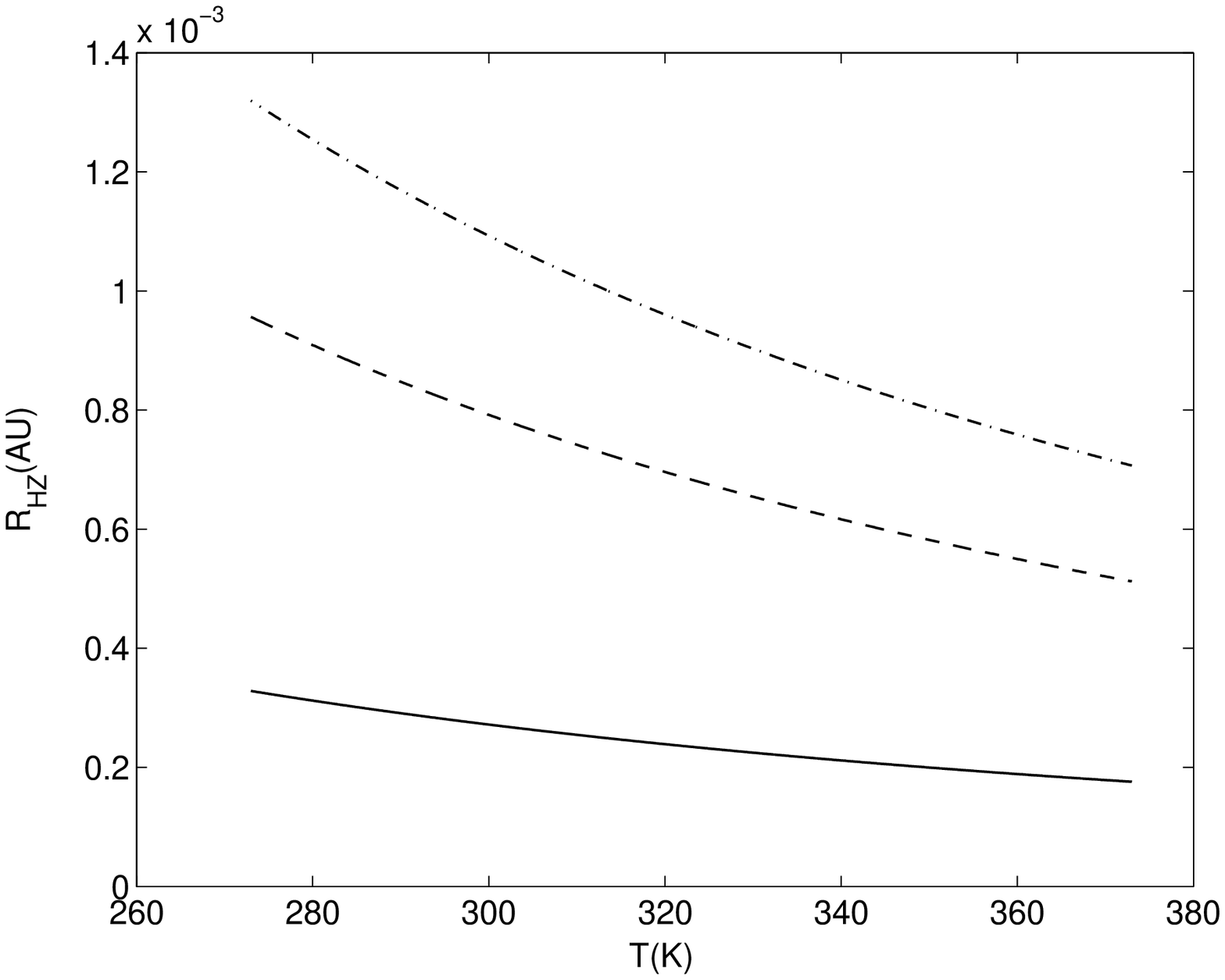}}
 \caption{On the top panel, in the framework of
the flux method, we plot the dependence of $T$ on $P$ for three
different values of $\dot{P}$: $\dot{P} = 10^{-15}$ss$^{-1}$ (solid
line); $\dot{P} = 10^{-14}$ss$^{-1}$ (dashed line); $\dot{P} =
2\times 10^{-14}$ss$^{-1}$ (dotted-dashed line). As it is clear from
the graph, for typical values of relatively slowly spinning pulsars
the effective temperature of the artificial construction ranges from
$\sim 400$K to $\sim 500$K. On the bottom panel, in the framework of
the temperature method, we show the dependence of $R_{HZ}$ on $T$
for the same values of $\dot{P}$. As we see the distance to the HZ
ranges from $2\times 10^{-4}$AU to $1.3\times 10^{-3}$AU.}
\label{fig2}
\end{figure}

After applying Eqs. (\ref{lumin},\ref{rHZ},\ref{angle}), one can
obtain $T$. On Figure \ref{fig2} (top panel) we plot the behaviour
of temperature versus the period of rotation of a pulsar for
different values of the slow down rate $\dot{P}$: $\dot{P} =
10^{-15}$ss$^{-1}$ (solid line); $\dot{P} = 10^{-14}$ss$^{-1}$
(dashed line); $\dot{P} = 2\times 10^{-14}$ss$^{-1}$ (dotted-dashed
line). As we see from the graph, for relatively slowly spinning
pulsars the typical values of the effective temperature of the
artificial construction are in the following interval $\sim
(400-500)$K. The corresponding radius of the HZ varies from $\sim
10^{-2}$AU to $\sim 10^{-1}$AU (see the bottom panel).

Generally speaking, the distance to the HZ is not strictly defined,
because our knowledge about life is very limited by the conditions
we know on Earth. But even in the framework of life on Earth the
radius of the HZ might be defined in a different way, assuming a
distance enabling the effective temperature in the interval:
$(273-373)$K where water can be in a liquid state (henceforth - the
temperature method) \footnote{In reality the temperature range might
be even narrower. For example the melting temperature of DNA is
approximately $60^o$C}. From Eq. (\ref{balance}) one can show that
\begin{equation}
\label{balance1} R_{HZ} =
\left(\frac{L}{8\pi\sigma\sin{\left[\beta/2\right]}T^4}\right)^{1/2}.
\end{equation}
On Figure \ref{fig2} (bottom panel) we plot the graphs of $R_{HZ}$
versus $T$ for the same values of $\dot{P}$. It is evident that in
this case the distance to the HZ ranges from $2\times 10^{-4}$AU to
$1.3\times 10^{-2}$AU.

Another possibility for the supercivilization might be the
"colonization" of millisecond pulsars since these objects reveal
extremely high values of luminosity. In particular, for the pulsar
with $P = 0.01$s and $\dot{P} = 10^{-13}$ss$^{-1}$ the slowdown
luminosity is of the order of $9.5\times 10^{38}$ergs s$^{-1}$ and
by taking into account $\kappa\approx 0.01$ (being quite common for
millisecond pulsars), one can show that the bolometric luminosity
$$L\approx 9.5\times 10^{36}\times
\left(\frac{P}{0.01s}\right)^{-3}\times\left(\frac{\kappa}{0.01}\right)
\times$$
\begin{equation}
\label{lb} \times\left(\frac{\dot{P}}{10^{-13}ss^{-1}}
\right)\times\left(\frac{M}{1.5M_{\odot}}\right)ergs\ s^{-1}
\end{equation}
is by several orders of magnitude higher than for relatively slowly
rotating pulsars. Here the physical quantities are normalized on
typical values of rapidly rotating neutron stars. It is clear that
all rapidly spinning pulsars have extremely high energy output in
the form of electromagnetic waves. As a result, a corresponding
distance from the central object to the HZ must be much bigger than
for slowly rotating pulsars.

In order to estimate the height of the "ring" one has to define the
opening angle of a radiation cone for millisecond pulsars radiating
in the $X$-ray spectrum. According to the classical paper
\citep{machus}, the $X$-ray emission of pulsars has the synchrotron
origin, maintained by the quasi-linear diffusion. As it was shown by
Machabeli \& Usov (1979) the corresponding angle of the radiation
cone is given by
\begin{equation}
\label{angle2} \beta\approx
8^o\times\left(\frac{0.01s}{P}\right)^{1/2}\times\left(\frac{R_{st}}{10^6cm}\right)^{1/2},
\end{equation}
which automatically gives an interval of height of the artificial
construction: $(0.02-0.15)$AU.

%\begin{figure}
%\resizebox{\hsize}{!}{\includegraphics{fig2a.eps}}
%\resizebox{\hsize}{!}{\includegraphics{fig2b.eps}}
% \caption{On the top panel we present the
%behaviour of $T$ versus $P$ in the framework of the flux method. It
%is evident that for rapidly spinning pulsars the effective
%temperature of the "ring" is in the following interval:
%$(540-660)$K. On the bottom panel, in the framework of the
%temperature method, we show the dependence of $R_{HZ}$ on $T$ for
%three different values of $\dot{P}$: $\dot{P} = 10^{-13}$ss$^{-1}$
%(solid line); $\dot{P} = 2\times 10^{-13}$ss$^{-1}$ (dashed line);
%$\dot{P} = 4\times 10^{-13}$ss$^{-1}$ (dashed-dotted line). As we
%see the distance to the HZ ranges from $10$AU to $30$AU.}
%\label{fig3}
%\end{figure}

\begin{figure}
\resizebox{\hsize}{!}{\includegraphics{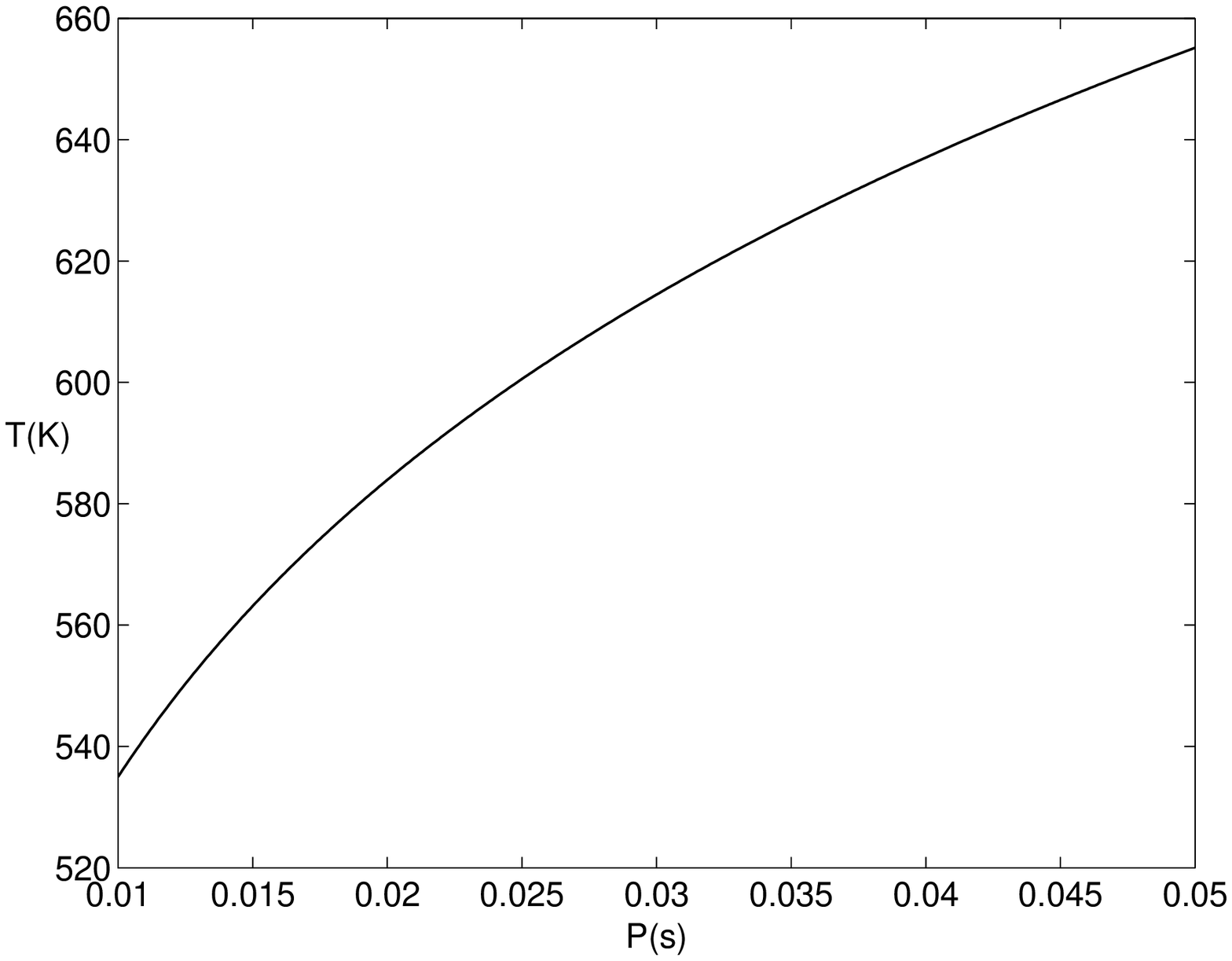}}
\resizebox{\hsize}{!}{\includegraphics{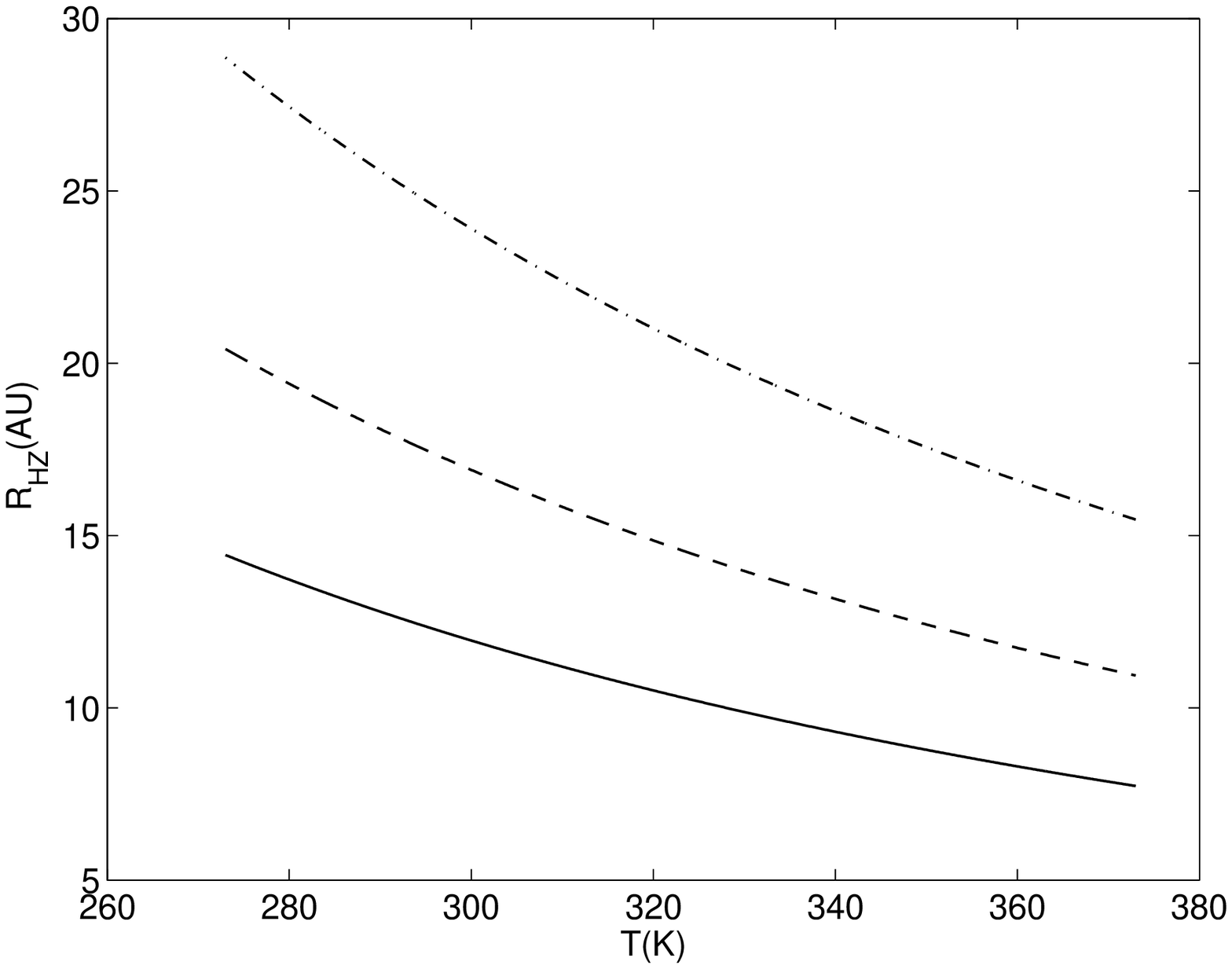}}
 \caption{On the top panel we present the
behaviour of $T$ versus $P$ in the framework of the flux method. It
is evident that for rapidly spinning pulsars the effective
temperature of the "ring" is in the following interval:
$(540-660)$K. On the bottom panel, in the framework of the
temperature method, we show the dependence of $R_{HZ}$ on $T$ for
three different values of $\dot{P}$: $\dot{P} = 10^{-13}$ss$^{-1}$
(solid line); $\dot{P} = 2\times 10^{-13}$ss$^{-1}$ (dashed line);
$\dot{P} = 4\times 10^{-13}$ss$^{-1}$ (dashed-dotted line). As we
see the distance to the HZ ranges from $10$AU to $30$AU.}
\label{fig3}
\end{figure}

After combining Eqs. (\ref{balance},\ref{angle2}), like the previous
case of slowly rotating neutron stars, one can estimate the
effective temperature for millisecond pulsars. In particular, on
Figure \ref{fig3} (top panel) in the framework of the flux method,
we show the behaviour of temperature of the "ring" as a function of
$P$ and as we see, for the following interval of rotation periods:
$(0.01-0.05)$s the temperature ranges from $\sim 540$K to
approximately $660$K. The size of the HZ ranges from $30$AU to
$350$AU. Contrary to this, by applying the temperature method, on
Figure \ref{fig3} (bottom panel) we show the behaviour of $R_{HZ}$
versus $T$ for three different values of $\dot{P}$: $\dot{P} =
10^{-13}$ss$^{-1}$ (solid line); $\dot{P} = 2\times
10^{-13}$ss$^{-1}$ (dashed line); $\dot{P} = 4\times
10^{-13}$ss$^{-1}$ (dotted-dashed line). From the figure it is clear
that the distance to the HZ ranges from $10$AU to $30$AU.

To understand how realistic are these constructions it is
interesting to estimate their masses. Let us assume that the
material the rings are made of has the density of the order of the
Earth, $\rho\sim 5.5$g cm$^{-3}$. Then it is straightforward to show
that the expression of mass is given by
\begin{equation}
\label{mass} M_{ring}\approx 4\pi\rho R_{_{HZ}}^2\Delta
R\sin\left(\beta/2\right),
\end{equation}
where $\Delta R$ is the average thickness of the ring. If we
consider slowly rotating pulsars, by assuming $\Delta R\sim 10$m and
$R_{_{HZ}}\sim 10^{-3}$AU (see Figure \ref{fig2}), one can show that
$M_{ring}\approx 4\times 10^{24}$g, that is three orders of
magnitude less than the mass of the Earth. Therefore, in the
planetary system the material could be quite enough to construct the
ring-like structure around $1$s pulsars.

The similar analysis for millisecond pulsars, shows that the mass of
the ring, $4\times 10^{32}$kg, exceeds the total mass of all
planets, planetoids, asteroids, comets, centaurs and interplanetary
dust in the solar system by three orders of magnitude. This means
that nearby regions of millisecond pulsars hardly can be considered
as attractive sites for colonization.

Another issue one has to address is the force acting on the
structure by means of the radiation. By assuming that half of the
total energy is radiated in each of the radiation cones, the
corresponding force can be estimated as follows
\begin{equation}
\label{fr} F_{rad}\approx\frac{L}{2c}.
\end{equation}
On the other hand, the stress forces in the ring, caused by
gravitational forces should be of the order of the gravitational
force acting on the mentioned area of the ring. This force is given
by
\begin{equation}
\label{fg} F_{g}\approx\frac{GM\lambda A}{R_{_{HZ}}^2},
\end{equation}
where $\lambda$ is the mass area density of the ring and $A = 2\pi
R_{_{HZ}}^2(1-\cos\left(\beta/2\right))$ is the area of the ring,
irradiated by the pulsar. It is clear that for maintaining stability
of the structure the radiation force should be small compared to the
gravitational force. By imposing this condition Eqs.
(\ref{fr},\ref{fg}) lead to $\lambda\gg L/\left(4\pi
GMc(1-\cos\left(\beta/2\right))\right)$, which for $\beta = 32^o$
reduces to
\begin{equation}
\label{sigma} \lambda\gg 3.4\times
10^{-6}\frac{L_{31}}{M_{1.5}}g\;cm^{-2},
\end{equation}
where $L_{31}\equiv 10^{31}$ergs s$^{-1}$ and $M_{1.5}\equiv
M/(1.5M_{\odot})$. As we see from the aforementioned estimate, for
realistic scenarios the emission cannot significantly perturb the
Dyson construction.

Generally speaking, rotation of pulsars drives high energy outflows
of plasma particles around the star \citep{guda} and the influence of these particles
deserves to be considered as well. The similar result can be obtained by
considering the interaction of the pulsar wind with the ring. In
particular, in Eq. (\ref{sigma}) instead of the radiation term,
$L/2$, one should apply the pulsar wind's kinematic luminosity, that
cannot exceed the slowdown luminosity. After considering the maximum
possible kinematic wind luminosity, $L_{sd}$, the critical surface
density will be of the same order of magnitude. Therefore, the Dyson
structure can survive such an extreme pulsar environment.

%\begin{figure}
%  \centering {\includegraphics[width=7cm]{fig4.eps}}
%  \caption{Here we show the in-plane displaced ring
%  with respect to the pulsar, denoted by $M$.}\label{fig4}
%\end{figure}

On the other hand, it is clear that gravitationally such a system
might not be stable. In particular, \cite{stability} has considered
a point mass and a thin solid ring having initially coincident
centers of masses, thus being in the equilibrium state. Although, it
has been shown that this equilibrium configuration is stable due to
the perturbations normal to the plane of the ring and is unstable
for perturbations within the mentioned plane.

\begin{figure}
  \centering {\includegraphics[width=7cm]{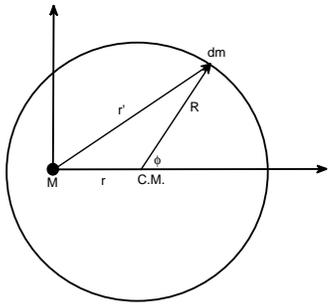}}
  \caption{Here we show the in-plane displaced ring
  with respect to the pulsar, denoted by $M$.}\label{fig4}
\end{figure}

On Fig. \ref{fig4} we schematically show the in-plane displaced ring
with respect to the pulsar. The gravitational force between the
pulsar and the ring element with mass $dm$
\begin{equation}
\label{grav1} df = G\frac{Mdm}{r'^2},
\end{equation}
projected on the $r$ line and integrated over the hole ring
\citep{stability}
\begin{equation}
\label{grav2} f_r(\zeta) = G\frac{MM_{ring}}{2\pi
R^2}\int_0^{2\pi}\frac{\zeta+\cos\phi}{\left(1+2\zeta\cos\phi+\zeta^2\right)^{3/2}}d\phi,
\end{equation}
for small values of $\zeta\equiv r/R$, leads to the equation
describing the in-plane dynamics \citep{stability}
\begin{equation}
\label{stab} \frac{d^2\zeta}{dt^2} -\frac{1}{\tau^2}\zeta = 0,
\end{equation}
where $\tau\equiv \sqrt{2R^3/(GM)}$ is the timescale of the process.
This equation has the following solution
\begin{equation}
\label{solut} \zeta(t) = \zeta_0 e^{t/\tau},
\end{equation}
indicating that the equilibrium configuration is unstable against
in-plane perturbations. Here, $\zeta_0$ is the initial
nondimensional perturbation. By means of this process the ring gains
the kinetic energy with the following power
\begin{equation}
\label{power} W = M_{ring}R^2\frac{d\zeta}{dt}\frac{d^2\zeta}{dt^2},
\end{equation}
therefore, in order to restore the equilibrium state of the ring,
one should utilize the same amount of energy from the pulsar
corresponding to the instability timescale. On the other hand, it is
evident that such constructions may have sense only if the power
needed to maintain stability is small compared to the bolometric
luminosity of the pulsar. By taking into account this fact, one can
straightforwardly show that the initial declination from the
equilibrium position must satisfy the condition
\begin{equation}
\label{cond} \zeta_0\ll
\frac{0.37}{R}\left(\frac{L_b}{M_{ring}}\right)^{1/2}\left(\frac{2R^3}{GM}\right)^{3/4},
\end{equation}
which for the parameters presented in Fig. \ref{fig3} leads to
$\zeta_0\ll\left( 10^{-5}-10^{-4}\right)$. It is worth noting that
such a precision of measurement of distance for supercivilization
cannot be a problem. For instance, in the Lunar laser ranging
experiment \footnote{The corresponding data is available from the Paris Observatory
Lunar Analysis Center: http://polac.obspm.fr/llrdatae.html} the distance is measured with the precision of the order
of $10^{-10}$.

In the vicinity of pulsars radiation protection could be one of the important challenges
facing the super advanced civilization. In particular, the radiation intensity
\begin{equation}
\label{int} I = \frac{L_b}{4\pi R^2\left(1-\cos\left(\beta/2\right)\right)},
\end{equation}
for the shortest ring radius $2\times 10^{-4}$AU is of the order of
$10^{12}$erg s$^{-1}$ cm$^{-2}$ and therefore, it should be of great importance
to protect the civilization from high energy gamma rays.
For this purpose one can use special shields made of certain material, efficiently
absorbing the radiation, which in turn, might significantly reduce the corresponding
intensity. If one uses half-value
layer, $HVL$, which is the thickness of the material at which the intensity is reduced by
one half we can estimate if the thickness of the ring is enough to make a strong protection
from extremely high level of emission. If as an example we examine concrete with $HVL=6.1$,
one can show that a layer of thickness $2.5$m reduces the intensity by $10^{12}$
orders of magnitude, being even more than enough for radiation protection.

%%%%%%%%%%%%%%%%%%%%%%%%%%%%%%%%%%%%%%%%%%%%%%%%%%%%%%%%%%%%%%%%%%%%%%%%%%%%%%%%
\section{Conclusion}
%%%%%%%%%%%%%%%%%%%%%%%%%%%%%%%%%%%%%%%%%%%%%%%%%%%%%%%%%%%%%%%%%%%%%%%%%%%%%%%%%%%%%

We extended the idea of Freeman Dyson about the search for
supercivilization and considered neutron stars. As a first example
we examined relatively slowly rotating pulsars, considering the
parameters $P=\left(0.5-2\right)$s; $\dot{P} = 10^{-15}$ss$^{-1}$;
$\dot{P} = 10^{-14}$ss$^{-1}$; $\dot{P} = 2\times
10^{-14}$ss$^{-1}$. It has been shown that size of the "ring" must
be by $(1-4)$ orders of magnitude less than those of the Dyson
sphere, which is thought to be of the order of $1$AU. The
corresponding temperatures of the artificial construction should be
in the following interval $(300-600)$K.

By considering the parameters of millisecond pulsars,
$P=\left(0.01-0.05\right)$s; $\dot{P} = 10^{-13}$ss$^{-1}$; $\dot{P}
= 2\times 10^{-13}$ss$^{-1}$; $\dot{P} = 4\times 10^{-13}$ss$^{-1}$
we found that the radius of the "ring" should be of the order of
$(10-350)$AU with an enormous mass $10^{32}$g exceeding the total
planetary mass (except the central star) by several orders of
magnitude. Therefore, it is clear that millisecond pulsars become
less interesting in the context of the search for extraterrestrial
superintelligence. Contrary to this class of objects, for slowly
rotating pulsars the corresponding masses of the Dyson structures
should be three orders of magnitude less than the Earth mass. We
have also examined the tidal stresses in terms of radiation and
pulsar winds and it has been shown that they will not significantly
perturb the Dyson construction located in the habitable zone if the
area density of the ring satisfies a quite realistic condition
$\lambda\gg 3.4\times 10^{-6}$g cm$^{-2}$. We have examined the stability
problem of the ring's in-plane dynamics and it has been shown that under certain
conditions the power required to restore the equilibrium position might be
much less than the power extracted from the pulsar. Also considering the problem
of radiation protection we have found that it is quite
realistic to reduce the high level of emission by many orders of magnitude.

It is worth noting that in the framework of the paper we do not
suggest that an advanced civilization would arise around a massive
star, surviving its supernova. On the contrary, we consider the
possibility to colonize the nearby regions of pulsars building
large-scale Dyson structures.

Generally speaking, the total luminosity budget of a pulsar is
emitted over a broad range of wavelengths, that can be harvested by
means of the Faraday's law of induction, transmitting
electromagnetic energy into that of electricity.

As we see, the pulsars seem to be attractive sites for super
advanced cosmic intelligence and therefore, the corresponding search
of relatively small ($0.0001$AU-$0.1$AU) infrared "rings" (with the
temperature interval $(300-600)$K) might be quite promising.

%%%%%%%%%%%%%%%%%%%%%%%%%%%%%%%%%%%%%%%%%%%%%%%%%%%%%%%%%%%%%%%%%%%%%%%%%%%%%%%%
\section*{Acknowledgments}
%%%%%%%%%%%%%%%%%%%%%%%%%%%%%%%%%%%%%%%%%%%%%%%%%%%%%%%%%%%%%%%%%%%%%%%%%%%%%%%%%%%%%

The research was partially supported by the Shota Rustaveli National
Science Foundation grant (N31/49).

\end{document}